\documentstyle[prl,epsf,psfig,graphics,twocolumn,aps]{revtex}
\begin{document}
\twocolumn[\hsize\textwidth\columnwidth\hsize\csname@twocolumnfalse%
\endcsname
\title{Energy radiation of moving cracks}
\author{S. Fratini$^1$,O. Pla$^1$, P.
Gonz\'alez$^2$, F. Guinea$^1$ and E. Louis$^3$
} \address{
$^1$ Instituto de Ciencia de Materiales, Consejo Superior de Investigaciones
Cient{\'\i}ficas, Cantoblanco, E-28049 Madrid, Spain. \\
$^2$ Departamento de F{\'\i}sica. Universidad Carlos III.
Butarque 15. Legan\'es. 28913 Madrid. Spain. \\
$^3$ Departamento de F{\'\i}sica Aplicada, Universidad de Alicante,
Apartado 99, E-03080 Alicante, Spain. \\}

\date{\today}

\maketitle

\begin{abstract}
The energy radiated by moving cracks in a discrete background
is analyzed. The energy flow through a given surface
is expressed in terms of a generalized Poynting vector.
The velocity of the crack is determined by 
the radiation by the crack tip.  The radiation becomes
more isotropic as the crack velocity approaches
the instability threshold. 
\end{abstract}

]
\narrowtext
\section{Introduction}
The dynamics of cracks in brittle materials are being extensively
studied\cite{Freund,FM99}, and a wealth of instabilities and patterns
have been observed as a function of control parameters such as
the applied 
strain\cite{Swinney,Ciliberto1,Ciliberto2,Fineberg1,Fineberg2}, 
or thermal gradients\cite{thermal}. The theoretical analysis
of moving cracks was initiated long ago\cite{Yo51,Sl81,KSS84}, with
the study of exact solutions for cracks moving at constant velocity.
These studies have been extended to a variety of different
situations\cite{Marder,KL99}. Alternatively, analytical approximations
to the leading instabilities of a moving tip have been 
proposed\cite{AB96}.

The simplest discrete model which captures the main features
of cracks in brittle materials is a a lattice with central forces
(springs) between nearest neighbors, whose bonds lose the restoring
force above a given threshold\cite{model1} (for
extensions see also\cite{model2}). This model, or simplifications
of it which leave out the vectorial nature of the strain field,
has been extensively used in modelling 
moving cracks\cite{FM99,KL99,Petal98,Petal00,Metal00}, although models
which deal with the microscopic structure of the system are also
being considered\cite{Abraham,Zetal97,Ketal97,Hetal99}.
Alternatively, various continuum models, which describe
the fractured zone in terms of additional fields,
have been proposed\cite{AKV00,KKL01}.

Discrete and continuum models of cracks differ in a variety of
features. It is known that the discrete models used so far cannot describe
a fracture zone at scales other than the size of the lattice cell in
the calculations\cite{Petal00,BP98}, although, even for a canonical
material such as PMMA, the fracture zone has a dimension much
larger than the size of its molecular building
blocks\cite{thesis}. 

Another important difference between
discrete and continuum models is the existence of radiation
from the tip of the moving crack, due to the existence of
periodic modulations in the velocity in the presence of
an underlying lattice. In this sense, a lattice model for
cracks is the simplest example where radiation due
to the scattering of elastic waves by deviations from
perfect homogeneity can be studied. These processes have been
observed in experiments\cite{Fineberg1,BC98,BC00,SCF01}, and it
has been argued that they are responsabible for 
some of the crack instabilities\cite{SCF01}. 

In the present work, we study the energy radiated by a crack
moving at constant velocity in a discrete lattice. We use a
generalization of the scheme discussed in\cite{Freund}.
The general method used is explained next. Section III presents
the main features of the results. The physical implications
of the results is discussed in section IV.

The problem of sound emission by moving cracks has been addressed,
within a different scheme, by\cite{BP01}. Insofar as the two approaches
can be compared, the results are compatible. Finally, radiation of 
moving cracks along the edge of the crack can be important in
understanding the roughness of the crack surface\cite{Betal01}. We will
focus on the radiation along the crack surface, and into the bulk of the
sample. Experiments\cite{SCF01} and simulations\cite{FM99}
suggest that this type of radiation
can play a role in the observed instabilities of the crack tip.
\section{The method}
We study discrete models of elastic lattices in two dimensional
stripes, as discussed in\cite{Petal98,Petal00}.
The underlying lattice is hexagonal, with nearest neighbor forces.
Bonds break when their elongation exceed a given threshold,
$u_{th}$, and under a constant strain at the edges, which,
scaled to the width of the stripe, we write as $u_0$.
We study models with and without dissipation in the dynamics
of the nodes.
Results depend on the ratio $u_{th} / u_0$ (see section III for further
details).
\subsection{Energy considerations}
In the absence of dissipation, the total kinetic plus elastic energy
must be conserved. In a continuum model, in the absence of
radiation, energy conservation leads to a global and to a local
constraint, for cracks moving at constant speed, $v$:

i) In the absence of radiation, the region well behind the crack tip
has relaxed to equilibrium, while the region ahead of it is
under the applied strain. The relaxed region grows at the expense
of the region under strain, at constant rate 
$\propto u_0^2 v W$ where $W$ is the
width of the stripe\cite{Petal00}. Energy is transferred to
the crack at this rate. As the energy stored in the crack
grows at rate $\propto u_{th}^2 v$, the crack can only
propagate (without radiation) for a fixed value of
$u_{th} / u_0$. Note that continuum solutions for radiationless
cracks moving at constant speed\cite{Yo51} do not specify a parameter
equivalent to $u_{th}$, so that they do not conflict with
energy conservation.

ii) The only position at which elastic energy is used to 
increase the size of the crack is the crack tip. Thus, at the
crack tip the flux of elastic energy should be equal
to the energy invested in enlarging the 
crack\cite{Freund,F72}. In continuum models, this local
constraint leads to an equation of the type:
\begin{equation}
\Gamma = A ( v ) G
\label{flux}
\end{equation}
where $\Gamma$ is the crack energy per unit length
($\Gamma \propto u_{th}$ in our lattice model), $G$ is
proportional to the stress intensity factor at the
crack tip ($G \propto u_0^2$), and $A ( v )$ is a universal
function which goes from 1 at $v=0$ to 0 at $v = v_R$ where
$v_R$ is the Rayleigh speed.

In lattice models, the energy arguments have to be modified
because of the presence of radiation of elastic waves. 
If we assume that the difference between continuum and
lattice models is small, we can use the perturbative scheme
discussed in\cite{WM95,LB95,RF97,MR98,MR00}. The crack tip
velocity undergoes oscillations at frequency 
$\omega = v / a$, where
$a$ is the lattice constant, and amplitude $f$. In order to 
estimate the energy radiated from the tip, we have to
extend the perturbative expansion to second order. 
We will not attempt here to calculate this expansion
rigorously. However, from the knowledge of the leading
term\cite{RF97,MR00}, we can infer that the radiation
due to a perturbation of frequency $\omega$ should go
as $B \omega^2$, where $B$ is a positive constant.
The global constraint i) implies that
$f^2 ( v / l )^2 \propto 1 - k ( u_{th} / u_0 )^2$, where
$l$ is the crack length and $k$ is a constant. 
The radiation also has to be taken
into account when balancing the energy absorbed at
the crack tip, ii) above. The corrections to eq. (\ref{flux})
lead to a condition of the type:
\begin{equation}
\frac{v}{v_R} = {\cal A} \left[ \frac{\left( \frac{u_{th}}{u_0}
\right)^2}{c + c' \left( 1 - k \left( \frac{u_{th}}{u_0}
\right)^2 \right)} \right]
\end{equation}
where $c$ and $c'$ are constants, and ${\cal A}$ is proportional to
the inverse of the function $A$ defined in eq.(\ref{flux}), the
magnitude in brackets being its argument.
Thus, the existence of radiation in discrete models
allows for the existence of a continuum of solutions
$v ( u_0 / u_{th} )$ for a given range 
of $u_0 / u_{th}$\cite{Marder,KL99,Petal98,Petal00,Metal00}.
Note that the existence of solutions which do not violate
energy conservation does not imply that these solutions
are stable. Full dynamical simulations of lattice
models\cite{Petal98,Petal00,Metal00} suggest that
inertial cracks (without dissipation) accelerate
until they reach speeds comparable to those predicted
by the Yoffe criterion\cite{Yo51}, and then bifurcate. 
\subsection{Energy flux: continuum elasticity}
In the following, we will reformulate the
concepts discussed in\cite{Freund} in order to 
make them more amenable for extensions to lattice models,
dicussed in the next subsection.

We describe an elastic medium in terms of the energy\cite{LL59}:
\begin{eqnarray}
{\cal H} &= &{\cal H}_{kin} + {\cal H}_{elastic} \nonumber \\
{\cal H}_{kin} &= &\int d^D r \frac{\rho}{2}
 \left( \frac{\partial {\bf \vec{u}}
( {\bf \vec{r}} )}{\partial t} \right)^2 \nonumber \\
{\cal H}_{elastic} &=  &\int d^D r \frac{\lambda}{2} \left(
\sum_i u_{ii} \right)^2 + \mu \int d^D r \sum_{ij}
u_{ij}^2
\end{eqnarray}
where $D$ is the spatial dimension, $\rho$ is the mass density, 
$\lambda$ and $\mu$ are Lam\'e
coefficients, ${\bf \vec{u}} ( {\bf \vec{r}} )$ denotes the
displacements at position ${\bf \vec{r}}$, and the $u_{ij}$'s
define the strain tensor:
\begin{equation}
u_{ij}  ( {\bf \vec{r}} ) = \frac{1}{2} \left( \frac{\partial 
u_i}{\partial r_j} + \frac{\partial u_j}{\partial r_i} \right)
\end{equation}
The equations of motion satisfied by ${\bf \vec{u}} ( {\bf \vec{r}} )$
can be written as:
\begin{equation}
\rho \frac{\partial^2 u_i}{\partial t^2} =
- \sum_j \frac{\partial}{\partial r_j} \sigma_{ji} 
\end{equation}
where $\sigma_{ji} = \partial {\cal H}_{elastic} / \partial u_{ij}$
is the stress tensor. 

The time derivative of the total energy $E_{\Omega}$ within a region
$\Omega$ is:
\begin{eqnarray}
\frac{\partial E_{\Omega}}{\partial t} &= &\frac{\partial}
{\partial t} \int_{\Omega}
d^D r \left [{\cal H}_{kin} + {\cal H}_{elastic} \right ] \nonumber \\
&= & \int_{\Omega} d^D r \left [ \rho \frac{\partial {\bf \vec{u}}}{\partial t}
\frac{\partial^2 {\bf \vec{u}}}{\partial t^2} +
\frac{\partial u_{ij}}{\partial t} \sigma_{ij} \right ] \nonumber \\
&= &- \int_{\Omega} d^D r \frac{\partial}{\partial r_j}
\left( \sigma_{ij} \frac{\partial u_i}{\partial t} \right)
\label{Omega}
\end{eqnarray}
so that the vector ${\bf \vec{P}} ( {\bf \vec{r}} )$ with
components $P_j = \sum_i \sigma_{ij} \partial u_i / \partial t$
plays the same role as the Poynting vector in electrodynamics.
The energy flux through an element of area $d {\bf \vec{S}}$
is given by ${\bf \vec{P}} d {\bf \vec{S}}$.
Note, however, that, unlike in electromagnetism, the equations
of elasticity have not Lorentz invariance (there are two
sound velocities), and it is not possible to define
a four vector combining ${\bf \vec{P}}$ and the energy
density. The energy transferred to the outside of this region
remains defined as the flux of the vector ${\bf \vec{P}}$
through the surface bounding $\Omega$.
In the presence of dissipation, we still use ${\bf \vec{P}}$
as defined in Eq. (\ref{Omega}) in the understanding that
what viscosity does is to trigger the partial absorbtion 
of the radiated enegy whithout
changing the direction in which it is emitted. 
The vector ${\bf \vec{P}}$ will be our starting point
in the study of the energy flux of a moving crack.

\subsection{Energy flux: lattice model}
We will compute numerically the radiation of energy in
a discrete model, defined as a hexagonal two dimensional lattice
with nearest neighbor forces\cite{model1,Petal98,Petal00}. 
The energy is given by the sum of a kinetic term, associated
to the velocities of the nodes, and an elastic term, due
to the deformation of the bonds. 
The variation of the elastic energy of a given bond
with time can be written as:
\begin{equation}
\frac{\partial E_{ij}}{\partial t} = k \left[ \left(
{\bf \vec{u}}_i - {\bf \vec{u}}_j \right)
{\bf \vec{n}}_{ij} \right]
\frac{\partial \left[ \left(
{\bf \vec{u}}_i - {\bf \vec{u}}_j \right)
{\bf \vec{n}}_{ij} \right]}{\partial t}
\end{equation}
where $k$ is the force constant, and
${\bf \vec{n}}_{ij}$ is a unit vector in the direction of the bond.
We distribute this energy among the two nodes connected
by the bond, so that we can write the total elastic energy
within a given region as a sum of the contributions of
the nodes within that region.
The variation in the kinetic energy at node $i$ is:
\begin{equation}
\frac{\partial K_i}{\partial t} = - k \sum_{j}
\frac{\partial ( {\bf \vec{u}}_i {\bf \vec{n}}_{ij} )}{\partial t}
\left[ \left(
{\bf \vec{u}}_i - {\bf \vec{u}}_j \right)
{\bf \vec{n}}_{ij} \right]
\end{equation}
The variation of the total energy within a given region 
is calculated by summing over all bonds within that region.
The kinetic and elastic contributions for all bonds outside the
edge of the region cancel. We are left with surface terms only,
as in the continuum model described earlier. The surface contributions
can be written as a sum of terms associated to the bonds which
connect a node within the region under study and a node outside.
Thus, a surface which includes a given node and has a given orientation
leads to an energy flux across it which can be calculated
from a weighted sum of
the positions and velocities of the bonds which connect
that node to its neighbors. As we can associate to each surface
orientation an energy flux, we can define the lattice
Poynting vector, in analogy to the analysis done for
the continuum model.  We will use this discrete Poynting vector
in the discussion of the energy dissipation of a moving crack
below.

\section{Results}

\begin{figure}[htbp]
\centerline{
\resizebox{8cm}{!}
{\rotatebox{-90}{\includegraphics{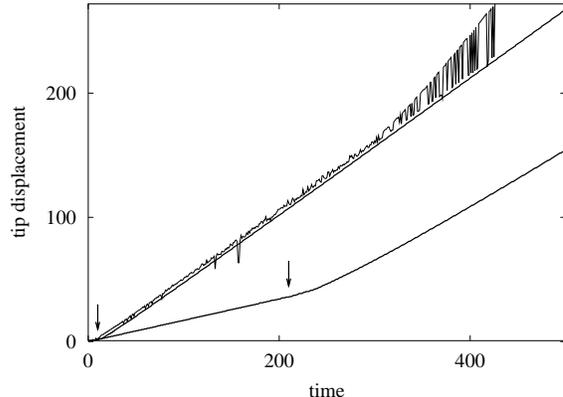}}}}
\caption{Crack tip displacement versus time for cracks under two different
applied strains (upper thick line: $u_0=0.08$; lower thick line: $u_0=0.02$) 
and zero viscosity. The arrow indicates the
position of the notch beyond which the stresses at the crack tip
exceed the threshold stress, and the crack propagates freely. The thin line is
for $u_0=0.08$, but allowing for the branching instability.}
\label{veloc}
\end{figure}

The discrete equations of motion in a two
dimensional lattice of a given size are integrated numerically
as discussed in detail elsewhere \cite{Petal00}.
The lattice is maintained under constant load at the edges.
In order to obtain cracks moving at constant velocities, 
a notch is induced at one side, which is gradually enlarged,
along a straight line, until the stress buildup leads
to the spontaneous propagation of the crack. 
The crack position, as function of time, is shown in Figure 
\ref{veloc}, for two different applied strains. The calculations
show that the crack propagates freely at a constant velocity in
the steady state. Our method for the calculation of the
properties of cracks moving at
constant speeds should lead to the same results as other 
techniques.

Instabilities are
avoided by allowing only the bonds directly ahead of the crack
to break. In other words, we force the crack to propagate straightly 
(with no branching). The simulations are performed in systems of 
$400\times 120$ lattice
sites, where we have checked that finite size effects on the steady 
state velocity are less than 1 percent. 

\subsection{Crack velocity}

\begin{figure}[htbp]
\centerline{
\resizebox{8cm}{!}{\rotatebox{-90}{\includegraphics{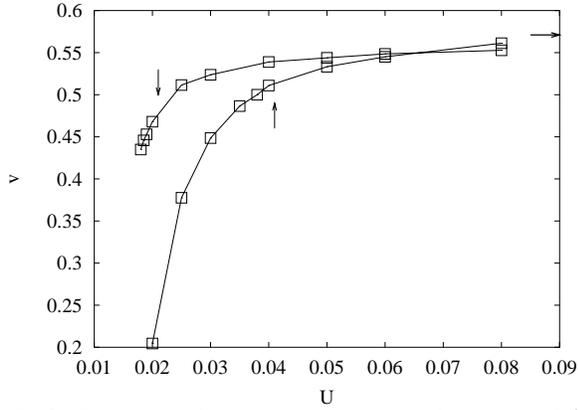}}}}
\caption{Crack velocity versus external strain, for $\eta=0$ (upper curve) 
and $\eta=0.8$ (lower curve). The arrow on the right  
indicates the Rayleigh velocity $v_R$. The vertical arrows mark the 
(avoided) branching 
instability (see text).
The threshold for breaking is 
$u_{th}=0.1$}
\label{v_vs_U}
\end{figure}

Figure \ref{v_vs_U} shows the steady state velocity $v$ as a function of the 
applied strain $u_0$, for two different values of the viscosity ($\eta=0$ and
$\eta=0.8$ in our units).  The crack velocity increases monotonically with $u_0$ and asymptotically tends to its limiting value $c_R=0.571$, the Rayleigh 
velocity in units where the force constant $k=1$ and the mass per site $m=1$ 
\cite{model2}.
Due to lattice trapping, there is a minimum allowed
$u_0$ whose value is roughly independent of $\eta$ \cite{KL99,trap},  
which in turn leads to a minimum crack speed which depends strongly on $\eta$.
The arrow marks the instability that would occur if the crack were not 
constrained to move on a straight line.

\subsection{Elastic energy and hoop stress}

\begin{figure}[htbp]
\centerline{
\resizebox{8cm}{!}{\includegraphics{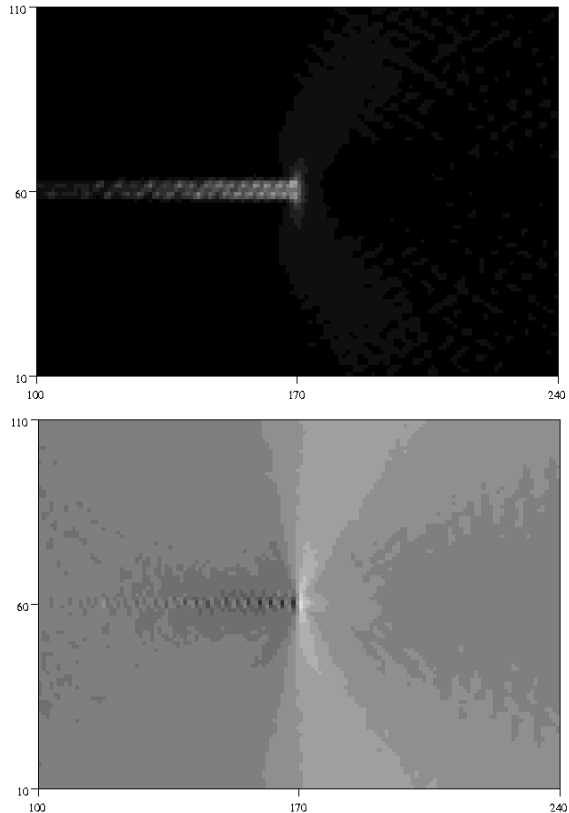}}}
\caption{Density of elastic energy (upper panel) and hoop stress
  (lower panel) for an inertial crack ($\eta=0$) moving 
under an applied strain $u_0=0.02$, below the branching instability.  
The crack moves from left to right, 
and the tip is located at a point of coordinates $170,60$.}
\label{low}
\end{figure}


Figure \ref{low} shows snapshots of 
the density of elastic energy and the 
hoop stress  at a given time $t_0$, 
for a steady inertial 
crack ($\eta=0$) moving at a 
velocity below the branching threshold (the applied strain is $u_0=0.02$, 
cf. figure \ref{v_vs_U}).

The density of elastic energy (Fig.\ref{low}, upper panel) 
has a sharp peak at the crack tip. 
In the near region (a few lattice spacings  
away from the tip), we see that the distribution of 
elastic energy is very anisotropic: it is sizeable in the direction 
perpendicular to the crack motion, where it decays smoothly with the distance, 
and all along the crack, where it has an oscillating behavior. This behavior
is reminiscent of the Rayleigh waves which propagate on the crack surface 
(see section \ref{radiation} below). At larger 
distances (of the order of the linear dimensions of the system), the elastic 
energy is smoother and has a broad maximum ahead of the tip, 
around a given angle of the order
of $\theta \approx \pi/3$ from the crack direction. 
We cannot be conclusive about this maximum being intrinsic in 
nature,  or rather being related to the symmetry of the underlying triangular 
lattice (see \cite{Metal00} for a more detailed discussion of this point).

The hoop stress (Fig. \ref{low}, lower panel) 
shows a very similar behavior, with  strong oscillations 
all along the crack, and maxima perpendicular to 
the crack motion, the maximum shifting from $\theta \approx \pi/2$ to 
$\theta \approx \pi/3$ with increasing  distance from the tip.

\begin{figure}[htbp]
\centerline{
\resizebox{8cm}{!}{\includegraphics{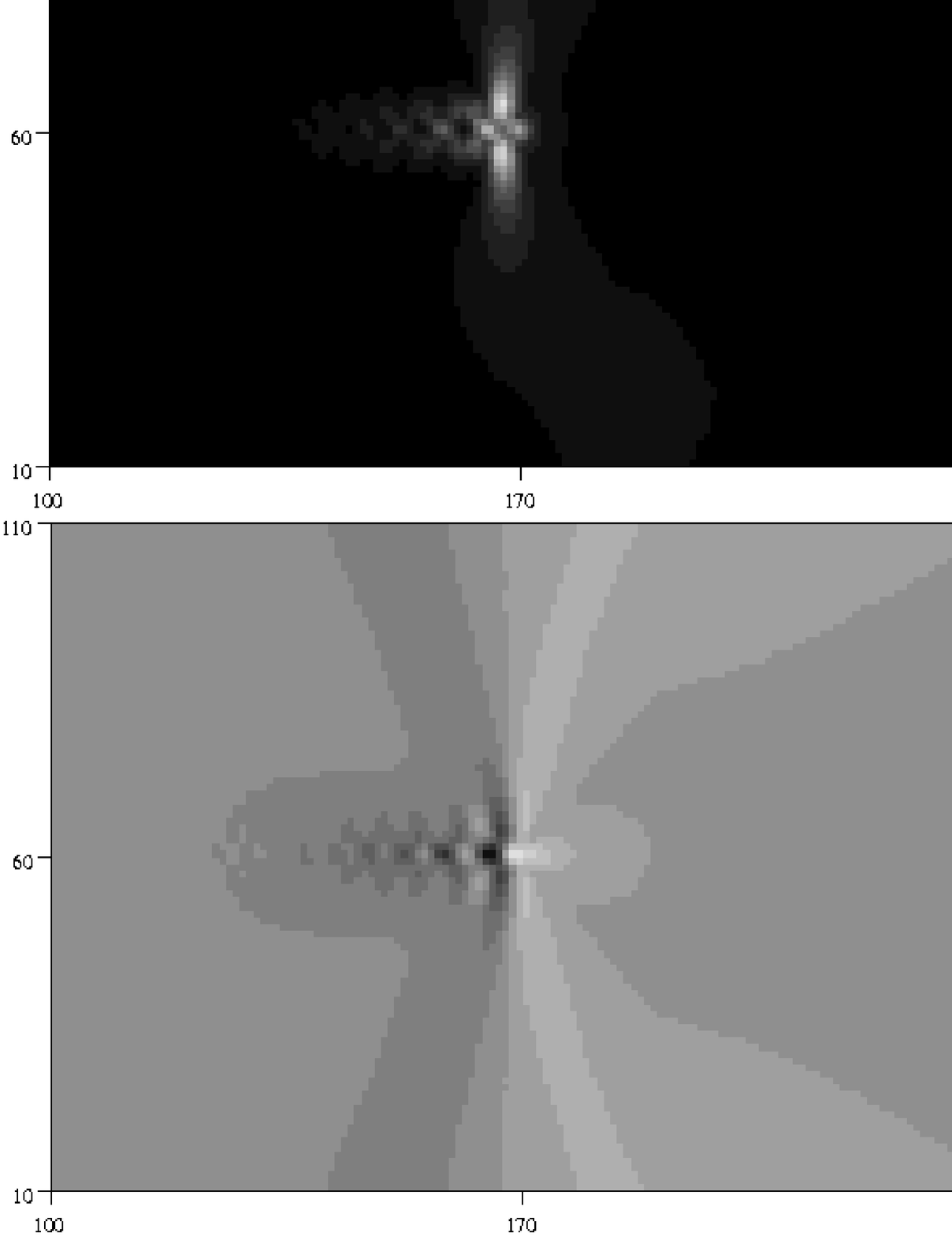}}}
\caption{Density of elastic energy (upper panel) and hoop stress
  (lower panel) for an inertial crack ($\eta=0$) moving 
under an applied strain $u_0=0.08$, well above the branching instability.}
\label{high}
\end{figure}


Figure \ref{high} is the same as 
Fig. \ref{low}, but  
for a crack moving at a 
velocity well above the branching threshold (the applied strain is $u_0=0.08$).
We notice that the distribution of elastic energy and  hoop stress 
has changed qualitatively: 
the bulk features in the direction
perpendicular to the crack motion now dominate over the oscillating part along
the crack. The latter decay more rapidly  and eventually disappear 
far behind the tip.

\begin{figure}[htbp]
\centerline{
\resizebox{8cm}{!}{\includegraphics{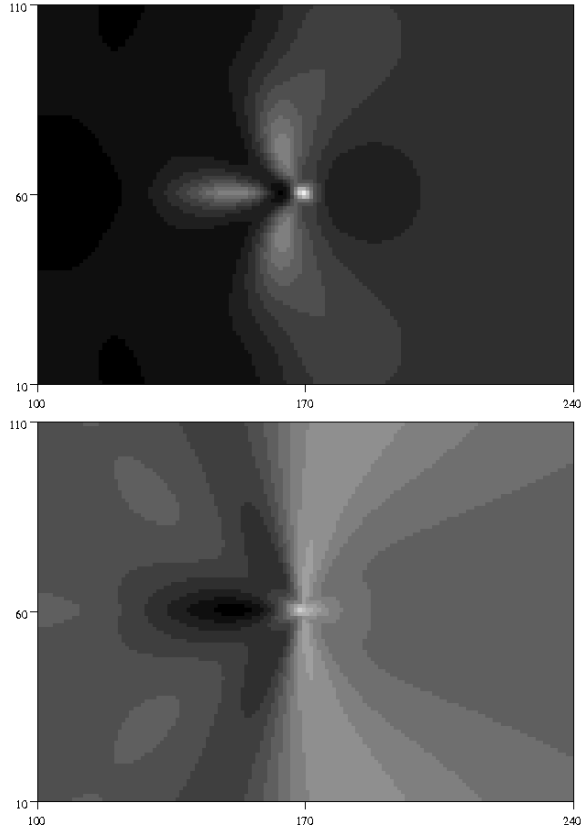}}}
\caption{Density of elastic energy (upper panel) and hoop stress
  (lower panel)  for a dissipative crack ($\eta=0.8$) moving 
under an applied strain $u_0=0.08$.}
\label{visco}
\end{figure}


The elastic energy and hoop stress corresponding to a 
\textit{dissipative}  
crack ($\eta=0.8$, $u_0=0.08$) 
are shown in fig. \ref{visco}.
Although the overall characteristics are similar to the inertial case, 
with maxima at the tip and in the direction transverse to the crack,
the distribution of stresses is much smoother. 
Moreover, the oscillations associated with 
Rayleigh waves along the crack are washed out by viscosity, being replaced
by a single broad maximum behind the tip.

\subsection{Radiation}
\label{radiation}

\begin{figure}[htbp]
\centerline{
\resizebox{8cm}{!}{\rotatebox{-90}{\includegraphics{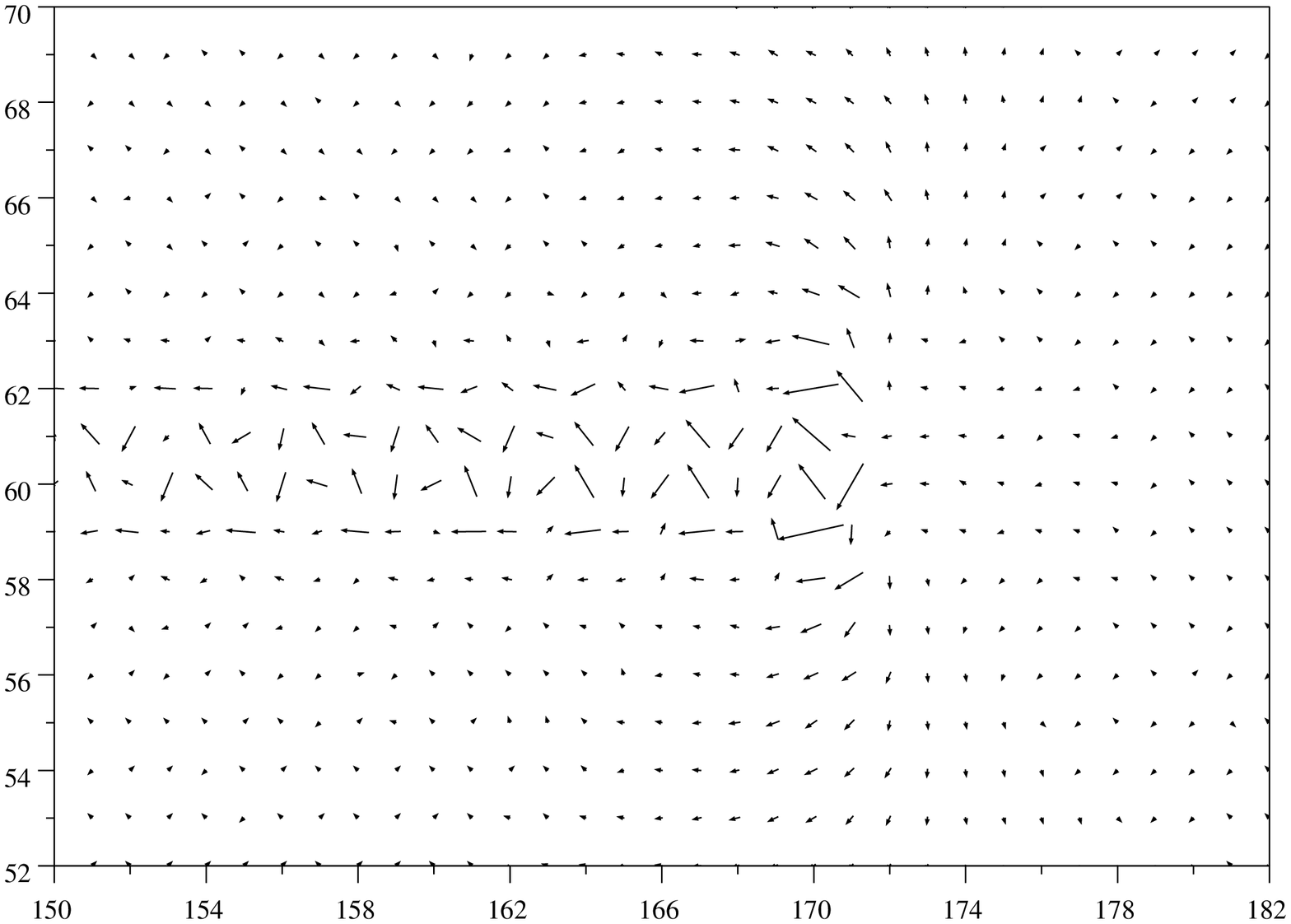}
\hspace{-3.2cm} \includegraphics{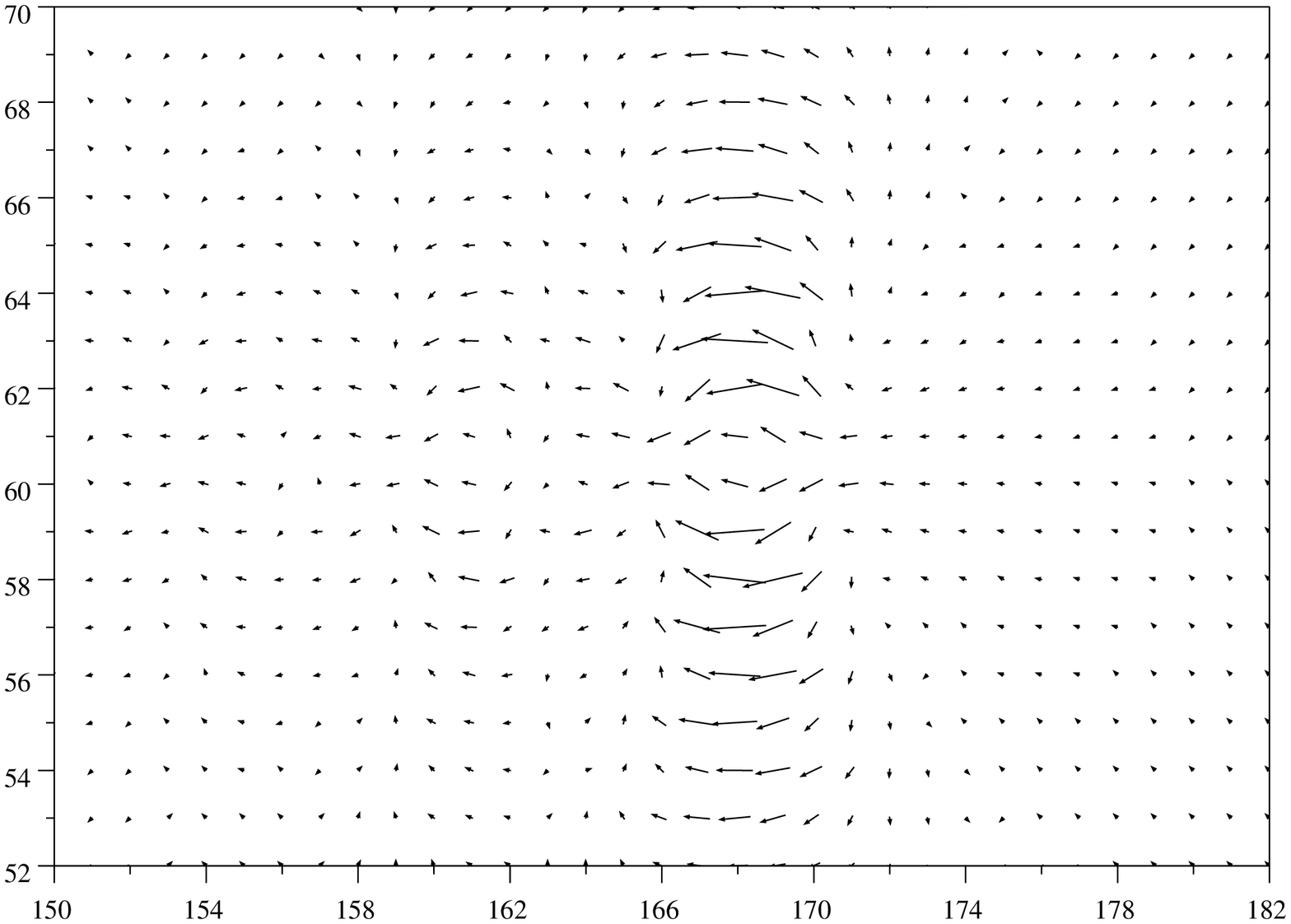}\hspace{-3.2cm}
\includegraphics{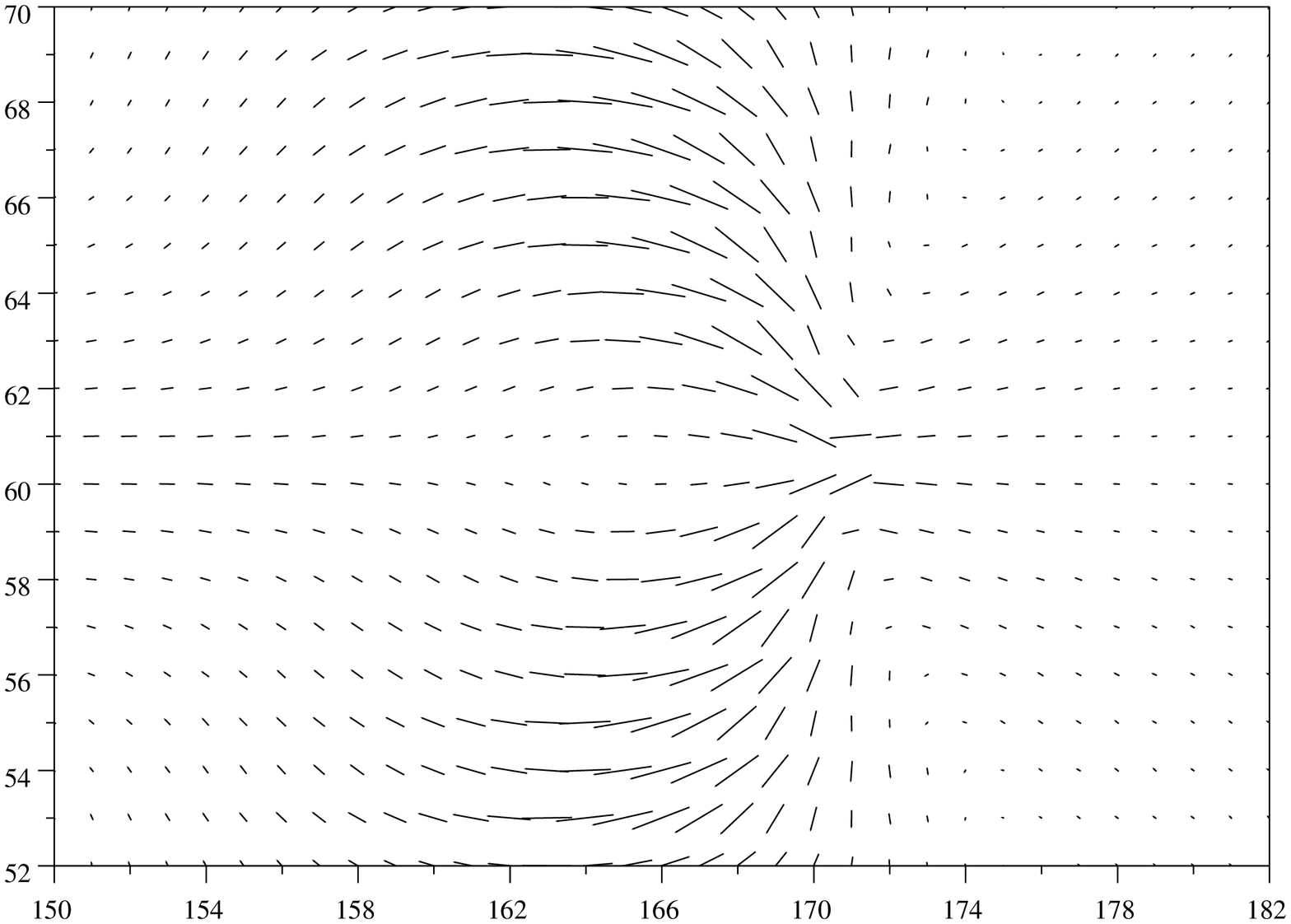}}}}
\caption{Poynting vector field representing the radiation propagating
  in the vicinity of the crack tip. Upper panel: slow inertial crack
  (same parameters as fig. \ref{low}); center panel: fast inertial
  crack (same as fig. \ref{high}); lower panel: dissipative crack
(same as fig. \ref{visco}).}
\label{vec}
\end{figure}

The above results can be better understood by analyzing  the  
Poynting vector field, which represents the flux of energy being radiated
at a given point in the system. 
As was stated in the introduction, emission of  sound waves 
is expected since the crack tip moves in a 
\textit{discrete}
medium, therefore acting as a source of radiation at a frequency
$\omega=v/a$,  the ratio of the crack speed to the lattice spacing. 
Moreover, one expects a net flux of energy in the direction 
\textit{opposite} to the crack motion, corresponding to the
elastic energy released from the region ahead of the tip, which 
allows the crack to move.

As can be seen in the first panel of figure \ref{vec}, at such moderate 
crack speeds most of the energy is radiated 
in the form of Rayleigh waves propagating backwards
along the crack, with a wavelength comparable 
with (but not equal to) the lattice spacing $a$. Despite the fact that
$\eta=0$, such waves are seen to decay at long distances behind the tip
(they decay into bulk waves, the oscillating bonds  on the 
crack surface acting themselves as sources of radiation). 
In addition, there is also
a weaker emission of bulk waves from the tip, 
responsible for the observed maximum
in the  direction perpendicular to the crack motion.

At high crack speeds, on the other hand (cf. center panel in 
fig. \ref{vec}), it
is the bulk radiation  which dominates the emission
pattern. Moreover, shadow images of the near-field appear behind the tip 
(the strongest one being at around $x=161$).  

In the case of viscous cracks (lower panel in figure \ref{vec}), the
emission pattern is entirely dominated by bulk waves, and Rayleigh
oscillations disappear in agreement with the reults of fig. \ref{visco}.

\begin{figure}[htbp]
\centerline{
\resizebox{8cm}{!}{\rotatebox{-90}{\includegraphics{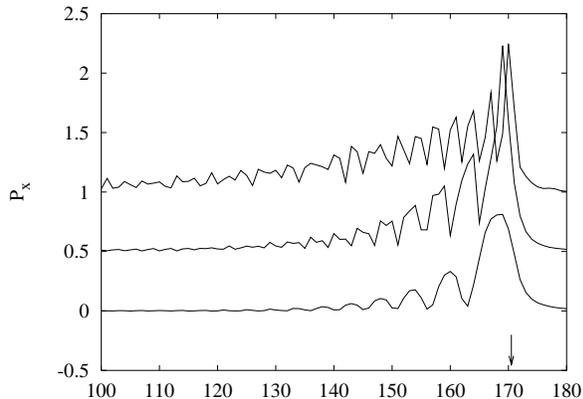}}}}
\caption{The component of the Poynting vector along an inertial crack, 
as a function of the coordinate $x$, normalized to the total energy flowing 
through the sample, $\propto u_0^2 v W$ (see text). From top to
bottom: $u_0=0.02, 0.04, 0.08$. The vertical arrow marks the tip position.
The curves are shifted by a 
vertical offset for clarity purposes.}
\label{rayleigh}
\end{figure}

In order to analyze the behavior of 
Rayleigh waves along inertial cracks, we plot in figure 
\ref{rayleigh} the component of the Poynting vector $P_x$ parallel to
the crack direction, at the surface of the crack, for different values 
of the applied strain. The data are normalized to
the difference in mechanical energy of a line far ahead from the crack, 
and a line far behind (this energy, which scales as $u_0^2vW$, 
is transferred to 
the crack in the fracture process). The figure clearly shows that 
the wavelength of surface waves as well as their decay rate   increase 
with the crack velocity.

\section{Conclusions}
We have analyzed the nature and influence of radiation in the
propagation of cracks in discrete systems. For the lattice
and force models that we have studied, we find:

i) Cracks in lattice models radiate energy, even when the
average velocity is constant and they move along a straight
line. This can be understood by assuming that the crack
tip undergoes oscillations at frequencies $n v / a$, where
$v$ is the velocity of the crack and $a$ is the
lattice constant.

ii) Radiation allows for the existence of a continuum of
solutions of moving cracks at constant velocity. 
The balance of static elastic and crack energy is
compensated by the radiation from the crack tip.

iii) At low velocities, most of the radiation is 
in Rayleigh waves along the surface of the crack. 
At velocities comparable to the  Rayleigh velocity,
a significant fraction of the radiated energy is
in bulk waves with a more isotropic distribution.

iv) Viscosity allows for a faster exchange of the
elastic energy stored ahead of the crack tip into 
other forms of energy. This can help to explain the
increased stability of straight cracks in the
presence of viscosity.

Among the questions which remain unsolved is the relation of
the radiation to the instabilities of the crack tip. 
Our results suggest that inertial cracks accelerate along
a straight line, until they attain speeds compatible
with Yoffe's criterion\cite{Yo51}. On the other hand,
the radiation of the crack tip becomes
more isotropic at high velocities. It is unclear
whether the continuum approach suffices to understand
the instability observed in dynamical simulations
of discrete models, or if the radiation from the tip of the crack
plays a role in the instability. Note that the
calculated instability occurs at higher velocities
than the instabilities observed experimentally.
\section{Acknowledgements}
We are thankful to R. Ball, P. Espa\~nol, M. Marder, T. Mart{\'\i}n,
A. Parisi, M. A. Rubio and I. Z\'u\~niga for helpful 
discussions. Financial support from grants PB96-0875 and
PB96-0085 (MEC, Spain), and FMRXCT980183 (European Union).

\end{document}